# Who Uses Whose Telescopes? Analyzing the Knowledge Geography and Research Dominance of Global Astronomical Facilities

Yue Chen[1], Yuqi Wang[1], Yuying Gao[1], Zhiqi Wang[1], Lianlian Liu[2], Chun Yang[1*]

**Abstract**

Large-scale research infrastructures (LSRIs) are central to contemporary science policy, combining massive capital investments with international access regimes. Yet whether open access to these infrastructures translates into more equitable scientific authority remains contested. Astronomy provides a critical case: world-leading observatories are globally shared but embedded in specific national contexts. We compile a novel country--year dataset (1955--2025) linking the location of astronomical facilities with publication usage and authorship roles. This enables us to distinguish between hosting, using, and leading in telescope-based research. Our analysis reveals: (i) usage and impact are heavily concentrated in a small number of facility hubs; (ii) scientific leadership is even more unequal than access or usage (Gini coefficient 0.91 for first/corresponding authorship versus 0.85 for facilities and usage); (iii) hosting and leadership often decouple--countries such as Chile and South Africa mediate large publication volumes without commensurate gains in leading roles; and (iv) global leadership has shifted from U.S. dominance to a multi-hub system centered in the United States, Western Europe, China, Japan, and Australia. These findings challenge the assumption that international access alone democratizes science. We argue that converting participation into leadership requires domestic PI programs, investments in instrumentation and data pipelines, and governance models that distribute credit more equitably. The study highlights how the governance of LSRIs shapes global scientific hierarchies and offers design principles for infrastructures that seek not only to share data but also to broaden scientific authority.

**Keywords:** Research infrastructures; Astronomy; Knowledge geography; Research leadership; Scientific inequality; International collaboration

Yue Chen, chenyuedlut@163.com; Yuqi Wang, wangyuqi2019@126.com; Yuying Gao, e1221758@u.nus.edu; Zhiqi Wang, zhiqi_wang90@126.com; Lianlian Liu, liulianlian@dlut.edu.cn; Chun Yang, yangchun@tsinghua.org.cn

* Corresponding author

1. Institute of Science of Science and S&T Management & WISE Lab, Dalian University of Technology, Dalian 116024, China; 2. School of Architecture & Fine Arts, Dalian University of Technology, Dalian 116024, China

# 1. Introduction

Large-scale research infrastructures (LSRIs) have become the backbone of twenty-first-century science policy. They concentrate capital, expertise, and advanced instrumentation; their governance—time allocation, data policies, authorship conventions—structured who can participate in frontier science and who can credibly claim intellectual leadership (Eggleton, 2024; Nocella et al., 2024; Rådberg & Löfsten, 2023). Astronomy is an archetypal LSRI field: flagship observatories demand exceptional sites and long investment horizons, mobilize transnational funding coalitions, and operate elaborate regimes for user access and data pipelines (Eggleton, 2024). As facilities have become more powerful and globally shared, a central policy question has sharpened: does international access democratize scientific authority, or does leadership remain concentrated despite shared infrastructure? Put simply: who uses whose telescopes—and who leads the science that results?

Existing literatures speak to fragments of this question. Work on LSRIs emphasizes the political economy of "big science" and the institutional persistence of flatworm governance (Baneke, 2020; Peters, 2006; Hallonsten, 2014). Studies in the geography of knowledge show uneven cross-national distributions of output and recognition (core–periphery patterns, "invisible colleges") (Gingrich, 2025; Lehuedé, 2023; Chen et al., 2020). Bibliometric research documents expanding international collaboration and the diffusion of participation via archives and open data ( Scarrà & Piccaluga, 2022; Börner et al., 2021). Yet scholarship on collaborative inequality suggests that authorship credit and agenda-setting roles remain more concentrated than participation, especially in large consortia ( Ma et al., 2015; Ellemers, 2021; Lehuedé, 2023). Astronomy exemplifies these dynamics: premium observing sites are geographically clustered (Atacama, Maunakea, the Canaries), data feed global teams, and bylines span many countries (Guridi et al., 2020; Lehuedé, 2023)-but leadership roles (first/corresponding authorship) may not mirror usage ( Ma et al., 2015; Ellemers, 2021).

Two gaps motivate our study. First, few analyses triangulate facility location, publications usage of those facilities, and authorship leadership within a single, harmonized framework. Second, it lacks a systematic, longitudinal view of how these relationships evolve as infrastructures and access regimes change. Without such evidence, policy debates about open access, capacity building, and credit allocation lean on case narratives rather than comparative indicators. We seek to narrow these gaps through a twofold approach. Conceptually, we connect the siting of research infrastructures with role-sensitive authorship indicators to show how platform governance structures scientific authority. Empirically, we document stylized facts about the global political economy of infrastructure and credit in astronomy, offering design implications for the governance of LSRIs beyond this field. Building on these insights, we pose four central research questions: How concentrated are facility stock, usage, impact, and leadership across countries? Do hosting and leading align, or do they decouple? How has leadership evolved—from a single dominant pole to a multi-hub system? And which governance arrangements and capability conditions enable the conversion of participation into leadership?

This paper assembles a country–year dataset (1955-2025) that links (i) a standardized registry of astronomical facilities, (ii) publications that explicitly acknowledge facility use, and (iii) national authorship roles and citation outcomes. Using natural-language processing to extract facility entities from acknowledgements and abstracts, we match 142,000 facility-using papers (1955–2025) to 533 facilities across 165 countries. This enables us provide descriptive, policy-relevant indicators: distributional profiles (Ginis) of stock, usage, impact, efficiency, and leadership; a leadership–participation map that positions countries by role; collaboration networks connecting hosts and users; and time-series views of rebalancing in leadership.

Our study contributes to three debates. First, we link infrastructure geography to authorship roles within a single, harmonized framework, addressing a common limitation in studies that treat facilities, publications, and leadership in isolation. Second, we provide distributional metrics (Ginis) across stock, usage, impact, efficiency, and leadership, and introduce a leadership–participation map that visualizes role differentiation. Third, we offer time-series evidence on rebalancing and decoupling, complementing static portraits in the literature on globalization of science. From a methodological perspective, we utilize facility entity extraction from acknowledgements and abstracts, enabling direct linkage of papers to infrastructures. Substantively, we show how openness in access can coexist with concentration in leadership, identifying governance levers—PI programs, instrumentation leadership, pipeline capacity, and credit-allocation rules—that translate participation into lead authorship. While our empirical domain is astronomy, the framework is portable to other LSRI ecosystems where similar governance and pipeline dynamics prevail.

The paper proceeds as follows. Section 2 reviews the literatures on LSRIs, knowledge geography, and collaborative inequality, motivating our measurement choices. Section 3 details data construction, entity extraction, and variable definitions. Section 4 presents result on global distribution and usage, the alignment (or misalignment) between ownership and leadership, collaboration network structure, and the evolution of leadership. Section 5 discusses mechanisms and policy levers, while Section 6 outlines implications for the design and evaluation of LSRIs beyond astronomy. Throughout, our aim is not to adjudicate scientific merit, but to illuminate how infrastructure geographies and governance arrangements shape who uses whose telescopes—and who leads the science built upon them.

## 2. Theoretical Background and Literature Review

This section situates our study at the intersection of the political economy of large-scale research infrastructures (LSRIs), the geography of scientific knowledge, and the sociology of collaboration and credit. We synthesize three strands: (i) how LSRIs structure access and authority; (ii) how knowledge geography has evolved from core–periphery hierarchies to networked asymmetries; and (iii) how authorship conventions and collaboration regimes allocate leadership. We then position astronomy as a revealing testbed and derive operational implications for measurement and analysis.

### 2.1 LSRIs as platforms: economics, governance, and strategic roles

LSRIs—observatories, accelerators, genome centers, and analogous platforms—are defined by high fixed costs, long investment horizons, and complex governance (Eggleton, 2020). Their economic logic differs from that of typical research projects: capacity is lumpy and indivisible; marginal access is relatively cheap once capacity exists; and the platform's value rises with complementary assets (instrument development, data pipelines, trained users). These characteristics make LSRIs both public goods (broadly usable scientific capacity) and club goods (access governed by membership, time allocation, and proposal success) (D'Ippolito & Rüling, 2019). Governance must reconcile openness with stewardship: merit-based access, long-term maintenance, upgrade cycles, and data policies that balance priority rights and fair reuse (Dawes, 2010).

Institutionally, LSRI governance exhibits persistence and path dependence ( Sydow,

Schreyögg, & Koch, 2009). Rules embedded early—about time allocation, embargoes, authorship conventions, or instrumentation contributions—anchor future distributions of opportunity and credit. In practice, the authority to set roadmaps, define instrument suites, and steward data pipelines often rests with a small set of funders, host institutions, and long-standing consortia (Eggleton, 2020). As a result, the formal rhetoric of openness can coexist with asymmetric control over critical bottlenecks (Sydow, Schreyögg, & Koch, 2009). Such asymmetries do not necessarily imply exclusion; rather, they create "soft monopolies" over the most consequential channels of scientific production—proposal success, pipeline access, and authorship leadership.

Finally, LSRIs operate in a geopolitical register. They serve as instruments of science diplomacy and soft power, anchoring alliances and regional integration while signaling technological capability (Wagner & Simon, 2023; Rungius & Flink, 2020, 2020). In this sense, LSRIs are not merely technical assets but strategic platforms whose governance choices reverberate through the global distribution of scientific authority.

## 2.2 Access regimes and asymmetries: openness in practice

Access to LSRIs is structured by a layered regime: (i) proposal selection and time allocation; (ii) data rights and embargo periods; (iii) pipeline and software access; and (iv) publication and authorship norms. Although these rules aim to reward scientific merit, their operation in practice channels leadership opportunities toward actors with the capacity to formulate compelling proposals, assemble teams, build or lead instruments, and process complex data (D'Ippolito & Rüling, 2019). Studies show that, in large-team collaborations, authorship conventions and reward systems exhibit marked inequalities: geographic location, institutional prestige, and access to resources shape how authorship is distributed, especially the allocation of leadership positions (e.g., first and corresponding authors) (Hoekman & Rake, 2024). Moreover, de facto control over data pipelines and review committees—reinforced by funding leverage and institutional influence—often concentrates key decision-making authority and scholarly impact (D'Ippolito & Rüling, 2019). In platform settings such as biobanks, the assignment of authorship and credit likewise faces challenges of fairness and transparency, underscoring that authorship should be allocated on the basis of actual contributions rather than default rules (Kleiderman et al., 2018).

Two mechanisms are especially relevant. First, instrument leadership often confers privileged authorship positions and agenda-setting power. Teams that design, build, or operate key subsystems accrue technical authority and informal "claim rights" on high-impact observations (González-Alcaide et al., 2017; Söderström, 2023). Second, data pipeline dominance matters: when calibration, reduction, and analysis are channeled through a small set of expert groups, those groups become the default hubs for collaboration and corresponding authorship ( Weilbacher et al., 2020; Chinchilla-Rodríguez et al., 2019). Thus, even when observational data are internationally accessible, leadership may remain skewed toward actors who control instruments and pipelines.

## 2.3 Knowledge geography: from core–periphery to networked asymmetry

The distribution of scientific capacity and recognition has long been understood as uneven. Early formulations—the "core–periphery" model—emphasize a small set of wealthy systems that dominate infrastructure, journals, and high-impact output (Avin et al., 2018; Zelnio, 2012). In this view, peripheral systems participate but seldom lead; an "invisible college" of elite institutions sets agendas and standards (Avin et al., 2018; Kwiek, 2021).

Contemporary scholarship refines this picture without overturning it. On the one hand, international collaboration has expanded, and digital repositories have reduced some barriers to participation. On the other hand, authority has not diversified proportionally: agenda setting, editorial influence, and resource control remain concentrated (Qin et al., 2022). Network perspectives reconcile these observations by highlighting both place and connectivity as determinants of capacity (Sandström et al., 2022; Gui et al., 2019). Proximity to infrastructure and the quality of one's network position (brokerage, centrality, access to platform governance) jointly shape who produces, and who leads, frontier research (Pepe et al., 2024).

Astronomy is paradigmatic. Premium observing sites are geographically concentrated (climate, altitude, radio quietness), yet usage is highly international through time allocation and archival data. This combination—localized capacity, global access—creates a setting where participation can spread without a commensurate diffusion of leadership. Understanding this decoupling requires looking beyond publication counts to the roles that countries and institutions occupy in collaborative production (Lehuedé, 2023; Chinchilla-Rodríguez et al., 2019).

## 2.4 Collaboration, authorship, and the allocation of leadership

Leadership is multi-dimensional—authorship roles, agenda-setting, resources, and gatekeeping—and each dimension has distinct observables. In large collaborations, authorship lists are long, but prestige and responsibility are hierarchically distributed: first authorship typically signals intellectual ownership of the analysis; corresponding authorship signals coordination, responsibility, and often PI status (González-Alcaide et al., 2017; Chinchilla-Rodríguez et al., 2019). In instrument-driven fields, instrument or survey builders can accrue durable authorship rights. Editorial, funding, and programmatic roles also matter but are harder to observe consistently across countries (Chinchilla-Rodríguez et al., 2019).

The literature on collaborative inequality documents systematic asymmetries. Scientists from lower-resource contexts frequently contribute labor, data collection, or local expertise, yet they are under-represented in lead authorship and editorial influence (Boampong et al., 2024; Koch et al., 2025). Several mechanisms are implicated: proposal success correlates with prior reputation; instrument leadership confers recurring advantages; and language, training, and pipeline access create cumulative advantages (Ma et al., 2015; Li et al., 2022). Importantly, such inequalities can persist despite formal openness, because openness expands participation faster than it redistributes agenda setting (Li et al., 2022).

Two implications follow for measurement. First, publication volume alone is insufficient: high usage may coexist with low leadership (Lapidow & Scudder, 2019; Patience et al., 2019). Second, role-sensitive indicators—first- and corresponding-author shares—are essential to capture the distribution of intellectual authority (Song, 2017).

## 2.5 Astronomy as an empirical testbed

Astronomy crystallizes the foregoing mechanisms in three ways. First, the field depends on site-specific, high-cost instruments (optical/IR telescopes at high, arid sites; radio arrays in radio-quiet zones; specialized particle-astrophysics and gravitational-wave instruments). This yields geographical concentration of capacity in a handful of corridors. Second, the community has strong traditions of international time allocation and archival data release, which materially broaden participation—especially for institutions without local facilities. Third, modern astronomy is pipeline-intensive: data calibration, reduction, and analysis often require sophisticated software ecosystems and human capital that are unevenly distributed. Together, these features produce a system in which participation can globalize while leadership remains anchored in infrastructure hubs and pipeline centers.

The field is also methodologically attractive. Facility usage is explicitly acknowledged in many publications (facility keywords, acknowledgements), enabling systematic matching of papers to infrastructures. Authorship roles are well coded in bibliometric databases. Finally, long time horizons permit temporal analysis of how participation and leadership evolve as new facilities come online and governance regimes change.

## 2.6 From theory to measurement: operational choices and limitations

Our conceptual framing implies a measurement strategy that aligns infrastructure siting, usage, and leadership roles at the country level. We operationalize:

**Infrastructure stock**: the number of facilities hosted by a country (Num_Facility). Stock is an imperfect proxy for capability—facilities differ in size and quality—but it enables cross-national comparisons and can be complemented with per-facility normalization.

**Facility-based usage**: the number of papers that explicitly acknowledge using facilities located in the country (Paper using Facility). This focuses on the realized output of domestic infrastructure, irrespective of authors' nationalities.

**Impact**: citation-based indicators for facility-using papers—h_index (cohort-year based) and citations per paper—to capture both volume-weighted and per-item influence.

**Efficiency**: paper_per_facility, the ratio of facility-using papers to hosted facilities, reflecting per-facility productivity (with caveats about heterogeneity in instrument scale).

**Leadership**: the sum of first- and corresponding-author papers (f+c) as a measure of leadership intensity, and shares of Author-f and Author-c to compare countries.

**Participation**: the share of papers on which a country appears as non-first, non-corresponding contributor (a proxy for usage without leadership).

These measures allow two complementary analyses. First, distributional profiles using Gini coefficients assess how concentrated stock, usage, impact, efficiency, and leadership are across countries. Second, a leadership–participation map (log–log) positions each country by its balance between leading and participating, with isoclines (e.g., y=x) to interpret relative roles. We further use collaboration networks to characterize structural positions (core–periphery, brokerage) consistent with prior literature.

Measurement has limitations that we acknowledge. Facility mentions under-count unacknowledged usage and may vary by journal norms; authorship conventions differ by subfield; and recent cohorts have shorter citation windows. Stock counts do not equal capacity; per-facility normalization is sensitive to small denominators. Our interpretation is therefore comparative and descriptive, designed to surface stylized facts rather than causal claims.

## 2.7 Expectations from the literature: stylized hypotheses

The theoretical and empirical literature suggests a set of expectations that guide our descriptive analysis:

E1: Top-heavy concentration. Stock and usage will be head-concentrated in a handful of infrastructure hubs (core–periphery logic; institutional persistence).

E2: Leadership supersedes usage in concentration. Leadership (first/corresponding roles) will be more concentrated than stock or usage, reflecting cumulative advantages in instrument leadership, proposal success, and pipeline control (collaborative inequality).

E3: Facility–leadership decoupling. Some host nations—especially those serving as global observing sites—will mediate high usage but capture a smaller share of leadership, given time-allocation rules and off-site PI concentration.

E4: Networked rebalancing. Over time, leadership will rebalance from a single dominant pole to a concentrated multi-hub system as new infrastructures come online and rising systems integrate into established consortia—without implying full democratization.

E5: Non-monotonic efficiency. Per-facility productivity will be highest for mid-scale, well-integrated portfolios and fall sharply where stock is very thin (scale economies, minimum efficient portfolios; denominator effects).

E6: Archives broaden participation more than leadership. Open data and archives will expand participation among non-host systems, but leadership will move more slowly unless domestic PI programs and pipeline capabilities co-evolve.

These expectations are not formal hypotheses to be tested causally; rather, they are benchmarks against which to read the patterns in our descriptive evidence.

## 3. Data and Methodology

### 3.1 Data sources and integration

We integrate three families of sources to link the geography of astronomical facilities to publication outcomes and authorship roles.

**(1) Facility registry**

We start from the American Astronomical Society (AAS) Facility Keywords registry (610 facilities as of June 2024). The registry standardizes facility names, keywords, locations (site/country), and facility types (e.g., optical/IR, radio/mm, high-energy, solar, particle/gravitational). This list provides the controlled vocabulary for recognizing facility mentions in publications and the authoritative mapping from facility to host country.

*Full Facility Name.* The official long form, often with aliases and operating organization(s). Example: China's Guo Shoujing Telescope lists "Chinese Academy of Sciences 4m Large Sky Area Multi-Object Fiber Spectroscopy Telescope (LAMOST; also known as Guo Shoujing Telescope)," capturing aliases such as "Large Sky Area Multi-Object Fiber Spectroscopy Telescope," "LAMOST," and "Guo Shoujing Telescope," and naming the operator.

*Keyword.* The canonical short label/acronym used in manuscripts (e.g., LAMOST).

*Facility Type.* Observing band(s) and class, with multi-label support to reflect composite capabilities. Bands include Gamma-Ray, X-Ray, Ultraviolet, Optical, Infrared, Millimeter, Radio, Neutrinos/Particles/Gravitational Waves, and Solar. Example: CHASE (Chinese H-alpha Solar Explorer) is labeled "Optical; Solar Facility," indicating cross-band capability.

*Location.* One of ten geographic partitions: Asia, Europe, North America, South America, Africa, Australia, Antarctica, Earth, Space, and Airborne. Example: LAMOST is labeled Asia.

**(2) Operational logs**

For major facilities with public records—e.g., ESO's VLT, ALMA, Subaru, Keck, FAST—we scrape or manually compile observation logs and time-allocation summaries (program IDs, semesters, partner shares, and PI affiliations). These logs are not used to construct usage variables (which are paper-based) but serve as face-validity checks on the direction and timing of facility usage inferred from publications.

**(3) Bibliometrics**

To comprehensively capture the astronomy literature while preserving metadata quality and workflow scalability, we rely primarily on Web of Science (WoS) and Scopus. Both platforms are long-standing, publisher-independent indexing services with broad journal and proceedings coverage, rapid update cycles, and rich, consistently structured fields (titles, abstracts, author affiliations, corresponding-author flags where available, acknowledgements/funding text, subject categories, and citation counts). Crucially for large-scale text mining and reproducibility, WoS and Scopus provide stable bulk export and programmatic access, enabling batch retrieval and longitudinal tracking with uniform schemas.

We treat the Astrophysics Data System (ADS) as a valuable complement rather than a primary source. ADS—maintained by the Harvard–Smithsonian Center for Astrophysics and NASA—offers excellent disciplinary breadth and convenient full-text linking for astronomy. However, for high-volume, fielded analyses, ADS is less suitable: (i) bulk harvesting is more constrained; (ii) several metadata elements critical to our pipeline (e.g., machine-readable acknowledgements/funding text, standardized corresponding-author flags) are less uniformly available; and (iii) field standardization can be heterogeneous across time and outlets.

Temporal scope. We assemble a long historical series (1955–2025) to contextualize build-out and impact waves. For text-mining reliability, the main analysis window is 1991–2024, when abstracts are systematically present, and acknowledgements coverage accelerates post-2008. All analyses that rely on acknowledgements (e.g., facility extraction) are flagged accordingly, and robustness checks use abstract-only matches for the pre-2008 period.

Unit of analysis. The primary unit is country–year (host or author-affiliation country), with paper-level data used to compute shares and role assignments. Countries are defined using a harmonized ISO mapping; historical name changes and territory codes are collapsed to present-day country entities (details in the appendix).

## 3.2 Corpus construction and harmonization

### 3.2.1 Constructing a Structured Lexicon of Astronomical Facilities

Astronomical facilities underpin large-scale data acquisition, experiments, and observations. As global astronomy has expanded, facilities are cited in heterogeneous ways across publications, which fragments records and impedes reliable aggregation. To standardize facility acknowledgment and indexing, the American Astronomical Society (AAS) Journals maintain the Facility Keywords registry(*Facility Keywords*, 2024), which provides a controlled vocabulary for facility usage in scholarly articles, enabling consistent identification, retrieval, and longitudinal analysis of facility-based research. To ensure uniqueness, global recognizability, and long-term traceability, the AAS—via Data Editors and the AASTeX maintenance team—applies a formal review process that is periodically updated with each list release.

Building on the authority and coverage of the AAS Facility Keywords, we construct a structured lexicon of astronomical facilities through data mining and multi-source integration. The lexicon organizes information along three dimensions—basic attributes, operating entity, and operating time—captured in eight core fields: Name mapping (canonical ↔ aliases), Facility type, Geographic location, Operating institution, Operating country, Start of operation, End of operation, Upgrade (major overhaul) date(s). The workflow is shown in Fig. 1.

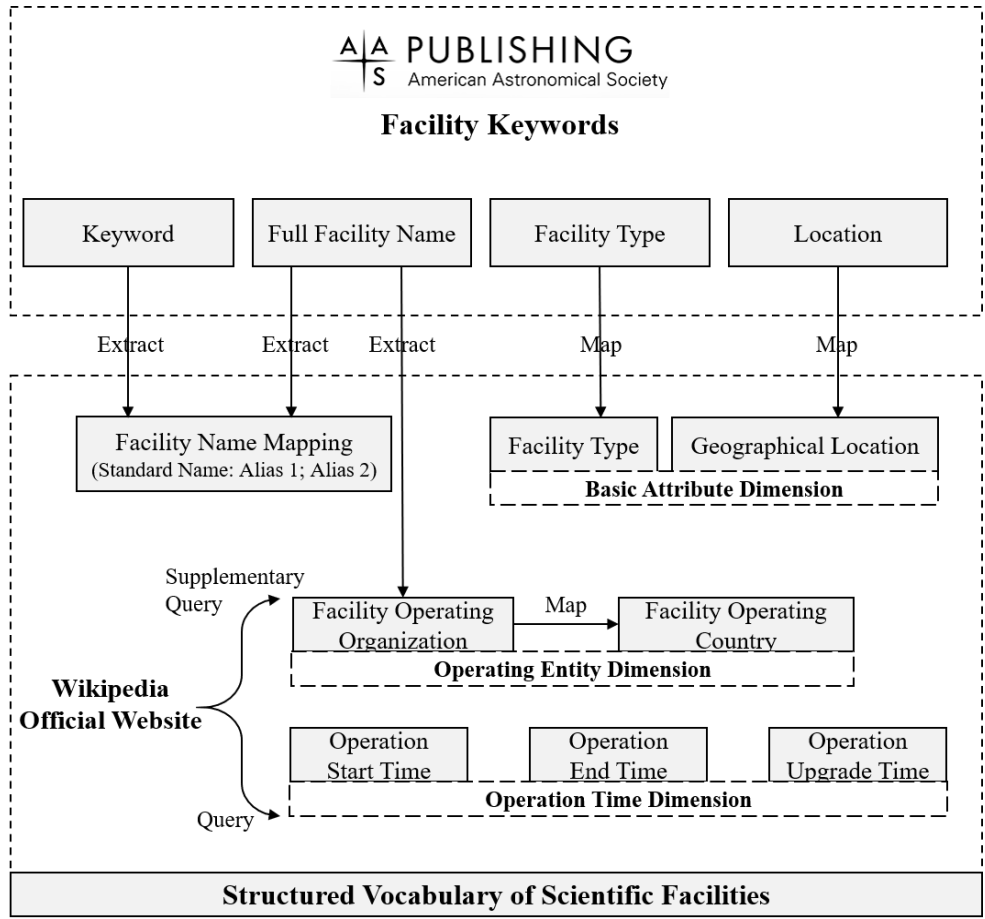

**Fig. 1** Overviews the construction process of the structured lexicon of astronomical facilities

**Step 1: Standardize facility names (name-mapping field).** Because a single facility may appear under multiple surface forms in the literature, we implement systematic alias extraction rules that parse parenthetical notes, acronym markers, and synonymous phrasing in the Full Facility Name, and merge these with the Keyword field. Human curation verifies precision and recall. We encode the result as "CanonicalName: alias1; alias2; …". Example: "LAMOST: Guo Shoujing Telescope; Large Sky Area Multi-Object Fiber Spectroscopy Telescope".

**Step 2: Extract basic attributes.** We directly ingest Facility Type and Location from the AAS registry to represent external attributes. These features are interrelated in practice—for example, radio arrays favor radio-quiet, remote sites; optical telescopes favor high, dry, dark locations (e.g., Andes, Maunakea); and gamma-ray observatories require extra-atmospheric platforms to avoid atmospheric absorption.

**Step 3: Extract operating-entity attributes.** Where present, the operating institution is parsed from Full Facility Name; otherwise, we consult authoritative sources (facility websites and Wikipedia) to complete coverage. We then construct an institution-to-country mapping to derive the operating country for each facility. Coverage reaches 100% of the 610 entries.

**Step 4: Extract operating-time attributes.** Because operational timelines are not in the AAS registry, we enrich with start of operations, end of operations, and major upgrade dates by harvesting from official facility histories and Wikipedia. After normalization and verification, these fields achieve 100% coverage across the lexicon. The resulting lexicon integrates three dimensions and eight fields for all 610 facilities, providing high-quality, analysis-ready data for downstream tasks including: (i) harmonized entity identification in acknowledgments/abstracts, (ii) stratified scientometric evaluations by band, type, and region, and (iii) lifecycle-aware analyses of usage, impact, and leadership over time.

### 3.2.2 Constructing the Astronomy Publication Dataset

To evaluate how scientific facilities underpin knowledge production in astronomy, we move beyond macro descriptions of facility types, geography, operating models, and historical evolution to link specific facilities to the scholarly outputs they support. The key challenge is to determine, at scale and with acceptable precision, which papers were enabled by which facilities. Harvesting publication lists from facility websites is infeasible (irregular updates, heterogeneous formats, missing or legacy sites). We therefore use the publications themselves as the evidentiary substrate and infer facility usage by mining paper text.

In astronomy, investigators typically acknowledge facilities explicitly in the Acknowledgements section when they use observing time, proprietary pipelines, or other core services. Acknowledgements are an institutionalized component of scholarly communication that record contributory organizations and individuals (e.g., funding agencies, observatories, survey teams). Prior research has leveraged acknowledgements to complement citation-based assessment, map collaboration and funding flows, and trace informal cooperation (Giles & Councill, 2004; Rigby & Julian, 2014; Rose & Georg, 2021); historical limitations were mainly data scarcity and formatting heterogeneity. Since 2008, Web of Science (WoS) has indexed acknowledgements, enabling large-scale analyses; Scopus has long captured a Funding Text field, though coverage varies by journal and year.

We draw on two authoritative bibliographic platforms, Web of Science (WoS) and Scopus to maximize coverage and metadata uniformity. Within subject categories of Astronomy & Astrophysics, we retrieved 726,261 records published 1945–2025 (downloaded 12 Apr 2025; "plain text" export) from WoS. WoS provides titles, abstracts, author keywords, authors and affiliations, corresponding-author flags (for most journals), acknowledgements, funding information, subject categories, and citation counts. Because Scopus lacks a dedicated field-level subject filter, we used ASJC 3103 (Astronomy and Astrophysics) to identify 105 relevant journals and retrieved 586,678 documents from 1855–2025 (downloaded 23 May 2025; "plain text" export). Scopus contributes complementary coverage and additional funding-text signals.

After deduplication (title-normalized keys + DOI/volume–page heuristics) and integration, we obtained a consolidated astronomy corpus (1885–2025) containing 937,701 unique records with titles, abstracts, author keywords, authors, affiliations, acknowledgements/funding text, and subject tags. Records were ingested into a relational database for structured storage and downstream text mining.

### 3.2.3 Facility-Entity Extraction and Matching

Because scientific facilities provide indispensable technical resources and services, their explicit recognition in scholarly outputs is a necessary precondition for linking infrastructure to knowledge production. We therefore extract—and disambiguate—named entities referring to astronomical facilities from two high-signal textual fields in the literature: Acknowledgements and Abstracts. The pipeline below (cf. Fig. 2) yields a reproducible map from paper→facility.

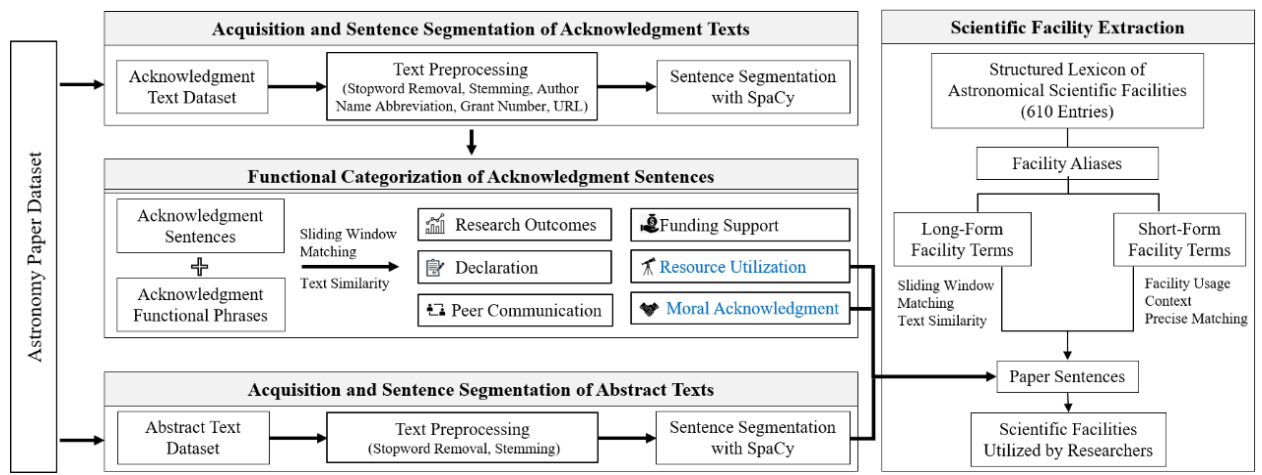

**Fig. 2** Entity extraction workflow for information on astronomical facilities

**(1) Text preparation and sentence segmentation**

We begin from an astronomy corpus of 937,701 records (1885–2025). For each record, we construct two text corpora, an Abstracts dataset and an Acknowledgements dataset (Funding/Acknowledgement text where available).

Both corpora are cleaned by (a) removing URLs, grant numbers, and disruptive abbreviations, (b) standard stop-word handling, and (c) Unicode and punctuation normalization. We then segment text into sentences using spacy en_core_web_trf (a Transformer/BERT-based model that improves boundary detection in technical prose). This produces 5,358,257 abstract sentences and 1,432,039 acknowledgement sentences as the atomic units for matching.

**(2) Functional classification of acknowledgement sentences**

To focus extraction on sentences most likely to carry facility mentions, we classify acknowledgement sentences into six functional categories derived from the contribution taxonomy in the acknowledgements literature reference to the existing research(Min Song et al., 2020): Financial, Resource use, Communication, Statement, Presentation, and General acknowledgement (Table 1).

**Table 1** Functions of Acknowledgement Sentences and Associated Noun Phrases

|   | Functions | Noun Phrases |
|---|---|---|
| 1 | Financial | finance; financial; funding; fund; studentship; award; fee; fellowship; invest; agency; grant; honoraria; honorarium; payment; salary; sponsor; foundation; scholarship |
| 2 | Resource use | make use of; operate by; provide; compute; facilities; facility; instrument; infrastructure; supporting acquisition; designed analysis; analysis software; analysis tools; bioinformatics analysis; chemical analysis; code; comment; develop; device; dna analysis; equipment; gene analysis; genetic analysis; hardware; histological analysis; image analysis; inspire; interpreting the analysis; laboratory analysis; laboratory tool; literature analysis; manage; mass spectrometry; meta-analysis; metabolite analysis; microarray analysis; molecular analysis; mrna footprinting; perform; phylogenetic analysis; proteomics analysis; resource; sample collection; server; spatial analysis; statistical analysis; structure analysis; support; technical; allocat; carried; carry; employ; employ; conduct; undertaken; acquisition data; analysis data; collect data; data access; data acquisition; data analysis; data collect; data from; data obtain; data process; data provide; data use; data were collected; obtain data; provide data; use data |
| 3 | Communication | study design; critical reading; referee; editorial view; editor; review; manuscript; advice; advise; discuss; edit; feedback; guidance; suggestion |
| 4 | Statement | conflicts of interest; do not reflect official; ethics approval; no direct role; no involvement; no involvement in; no role in; non fund; not assist with; not have any role in; not involved in; not necessary; official duty; views expressed are; responsible for; approval; authorization; compliance; |

| | | consent; copyright; declare; dedicate; disclaimer; disclosure; ethic; license; opinion; policy; www |
|---|---|---|
| 5 | Presentation | results presented; conference; poster; presentation; seminar; symposium |
| 6 | General acknowledgement | acknowledge; appreciate; thank; grateful; gratitude; recognize; indebt |

Classification uses fuzzy string matching over sliding windows combined with Levenshtein distance (edit distance) to allow for orthographic variation while controlling false positives. A sentence is assigned to a function if its best-match similarity to that function's lexicon exceeds a calibrated threshold.

Applied to 1,432,039 acknowledgement sentences, the distribution is: technical support 37.45%, Financial 24.36%, Communication 17.57%, General acknowledgement 14.36%, Statements 0.45%, Presentation 0.18%. Mentions of scientific facility use typically appear in acknowledgement sentences classified as "resource use" or "General Acknowledgement."

**(3) Facility Entity Extraction**

We construct a Structured Facility Lexicon from the AAS Facility Keywords registry (610 facilities), expanded with 2,169 aliases (canonical names, acronyms, historical names, multilingual variants). To manage ambiguity, aliases are partitioned into: Long tokens (≥10 characters or semantically distinctive strings): 434 items; Short tokens (<10 characters and prone to ambiguity): 1,735 items—e.g., HST, MOST, ISO, VISTA, FAST, where clashes with people's names, common words, or non-astronomy acronyms are common.

For long aliases, we apply sliding-window + Levenshtein matching over (a) Abstract sentences and (b) Acknowledgements sentences classified as Resource Use or General Acknowledgement (the two functions most likely to carry formal facility credit). Matches are retained if similarity exceeds a conservative threshold and if they co-occur with usage contexts (e.g., observed at/with, based on observations from, data obtained at, supported by instrument X).

For short aliases, we require exact boundary matches (token-level) and the presence of a recognized facility-use context. To operationalize context, we compile 124 regular-expression rules capturing assertion patterns (Table 2).

**Table 2** Selected Contextual Rules for Scientific Facility Usage

| No. | | Contextual Rules for the Use of Scientific Facility |
|---|---|---|
| 1 | Rule: | Facility: {fac}, {fac}, {fac} |
| | Example: | Facility: **HST**, **Spitzer**, **Fermi**. |
| 2 | Rule: | observations.*(made\|obtained\|taken)? .* (with\|using\|by) {fac} |
| | Example: | Observations were obtained with **ALMA**. |
| 3 | Rule: | (uses\|used\|makes use of\|made use of)? (data\|observations\|images)? (from\|provided by\|made by) {fac} |
| | Example: | This work makes use of data from the **LAMOST**. |
| 4 | Rule: | (data\|image\|spectra?) .* (was\|were) .* (provided\|produced\|enabled) {fac} |
| | Example: | Spectra were provided by **Keck**. |
| 5 | Rule: | (data\|image\|spectra?) .* (from\|using) {fac} telescope |
| | Example: | Data from the **Subaru** telescope were analyzed. |

When both channels (Acknowledgements and Abstract) produce matches, Acknowledgements take precedence (normative credit), with Abstract matches used to supplement recall. Multiple facilities may be associated with a single paper; we retain all.

Each retained match is mapped to facility → host country and attached to the paper ID. At the country-year level we compute: Paper using Facility (inclusive counting: a paper increments each host country whose facility it acknowledges); per-facility productivity (Paper using Facility / Num_Facility, for countries with ⩾ 1 facility);leadership and participation metrics (Author-f / Author-c shares; non-lead usage shares) computed from authorship affiliations. This linkage underpins all stock-usage-impact-leadership indicators in the Results.

**(4) Quality assurance, error profile, and limitations**

Stratified manual audits (by year, journal, facility type) indicate high precision in the Acknowledgements channel; Abstracts increase recall. Because thresholds and context rules are conservative, residual error biases toward under-counting (i.e., our usage estimates are lower bounds). Edge cases (MOST, VISTA, IRIS, ISO, FAST, WIND) are handled by negative lists and context constraints; examples of true positives include HST, Spitzer, Fermi, ALMA, LAMOST Acknowledgement coverage improves after 2008; pre-2008 identification leans more on Abstracts. Corresponding-author tags are not uniformly coded across outlets; we therefore privilege first authorship as the primary leadership proxy. Per-facility productivity is sensitive when Num_Facility is very small; we flag such cases in figures and avoid policy inference from local spikes.

Applying the above yields 142,445 facility-using papers linked to 533 facilities and 165 countries (main window 1991–2024; historical extension to 1955 for context). The method provides a transparent, scalable basis for measuring where facilities are hosted, which papers used them, and how usage relates to impact and leadership in the global astronomy system. The matched set underpins all stock–usage–impact–leadership indicators reported in Section 4.

### 3.3 Variable definitions and measurement

We operationalize constructs at the country–year level, distinguishing host country (by facility location) from authorship countries (by affiliations on each paper).

**(1) Stock (facility)**

**Num_Facility**: count of facilities physically hosted by country c. This is a stock measure, not weighted by aperture or budget. To partially address heterogeneity, we also report paper_per_facility (below).

**(2) Usage (facility-based output)**

**Paper using Facility**: count of publications in year t that acknowledge any facility hosted in country c. These measures realized output attributable to domestic infrastructure, regardless of where authors are based. A paper that uses two facilities in the same country counts once for that country; if it uses facilities from two different countries, each country's usage increments by one (inclusive counting). Sensitivity checks consider fractional allocation.

**(3) Impact (citations).**

**h_index (cohort-year)**: for papers in year t that use country c's facilities, we compute the h-index within that cohort (i.e., the largest h such that at least h papers from that cohort received $\geq$ h citations by harvest time). This combines volume and influence while limiting legacy accumulation.

**Citations per paper**: mean citations for that cohort.

**(4) Efficiency (per-facility productivity)**

**paper_per_facility**: Paper using Facility divided by Num_facility (defined for countries with $\geq$1 facility). This crude rate normalizes usage by stock and highlights mid-scale systems that convert compact portfolios into output. The measure is sensitive to small denominators.

**(5) Leadership (authorship roles)**

**Author-f (count/share)**: number (and share) of papers in which at least one first author's affiliation is in country c. When multiple affiliations exist for the first author, we attribute to all listed countries (inclusive).

**Author-c (count/share)**: analogous for corresponding author. Where the corresponding-author tag is missing, the paper drops from the Author-c share denominator; we do not impute.

**f+c**: sum of Author-f and Author-c counts as a leadership intensity index (recognizing overlap; the index is a simple additive proxy capturing opportunities to lead or coordinate).

**(6) Participation (non-lead presence)**

**Usage share (non-lead)**: share of papers where country c appears on the byline only as a non–first, non–corresponding affiliation. This complements Author-f/Author-c by capturing participation without lead roles.

## 3.4 Analytical strategy

We organize the analysis around four descriptive lenses.

**Distributional statistics.** We compute Gini coefficients for cross-country distributions of Num_Facility, Paper using Facility, h_index, Citations per paper, paper_per_facility, Author-f, and Author-c. Ginis are computed on non-negative country vectors within the study window; when series contain many zeros (common for leadership), we report the Gini with zeros included and provide robustness where small-N outliers could dominate.

**Leadership–participation map.** We position countries in a log–log plane, with the x-axis = log(Author-f) and y-axis = log(Usage share, non-lead). We overlay reference isoclines (y=x, y=2x, y=0.75x) to interpret the ratio of participation to leadership. Symbol/label encoding indicates facility presence (none; 1–9; ⩾10 facilities). To avoid log(0), countries with zeros on either axis are not plotted in the baseline panel; an alternative panel adds +1 before logging as a robustness view (appendix).

**International collaboration networks.** We build two country-level networks: (i) a co-authorship network where nodes are countries and edges are weighted by the number of joint papers; (ii) an operational collaboration network linking host countries to author countries based on facility-using papers. We compute degree, betweenness, and eigenvector centrality; detect communities using Leiden modularity; and visualize weighted backbones to emphasize salient ties. The goal is to situate leadership and participation within structural positions (core, broker, periphery).

**Temporal dynamics.** We construct country-year series for (i) host shares of facility-using papers; (ii) Author-f and Author-c shares; and (iii) yearly h-index of facility-using cohorts. We report linear fits for long-run tendencies (slopes with standard errors), mindful that early periods are small-N and recent years suffer citation truncation.

## 3.5 Reproducibility, ethics, and data availability

All steps-from facility dictionary construction and text parsing to indicator computation—are scripted. Upon acceptance, we will release: (a) the facility dictionary (with canonical names, aliases, and country mapping); (b) code for entity extraction and country harmonization; and (c) country–year indicator tables (stock, usage, impact, efficiency, leadership, participation), subject to publisher data-sharing constraints. We do not release full-text records from proprietary databases. No human subjects were involved; analyses operate on published metadata and acknowledgements.

In combination, these data and methods allow us to triangulate where facilities are, who uses them, and who leads the resulting science. The design privileges transparency over causal identification and emphasizes comparative magnitude and structural pattern. The subsequent Results section reports (i) distributional inequalities across stock, usage, impact, efficiency, and leadership; (ii) the leadership–participation landscape with facility-presence encoding; (iii) collaboration structure; and (iv) temporal rebalancing of hosting, usage, and leadership.

# 4. Results

## 4.1 Global distribution and usage of astronomical facilities

We observe a pronounced hemispheric asymmetry in facility siting: the Northern Hemisphere hosts the majority of installations (Europe, North America, East Asia), while several of the world's most intensively used ground-based sites lie in the Southern Hemisphere (notably Chile, South Africa, and Australia). This leads to a resource concentration vs. usage diversification pattern: infrastructure is geographically concentrated, yet usage is highly international as time-allocation programs, archival access, and multinational consortia distribute observing opportunities widely.

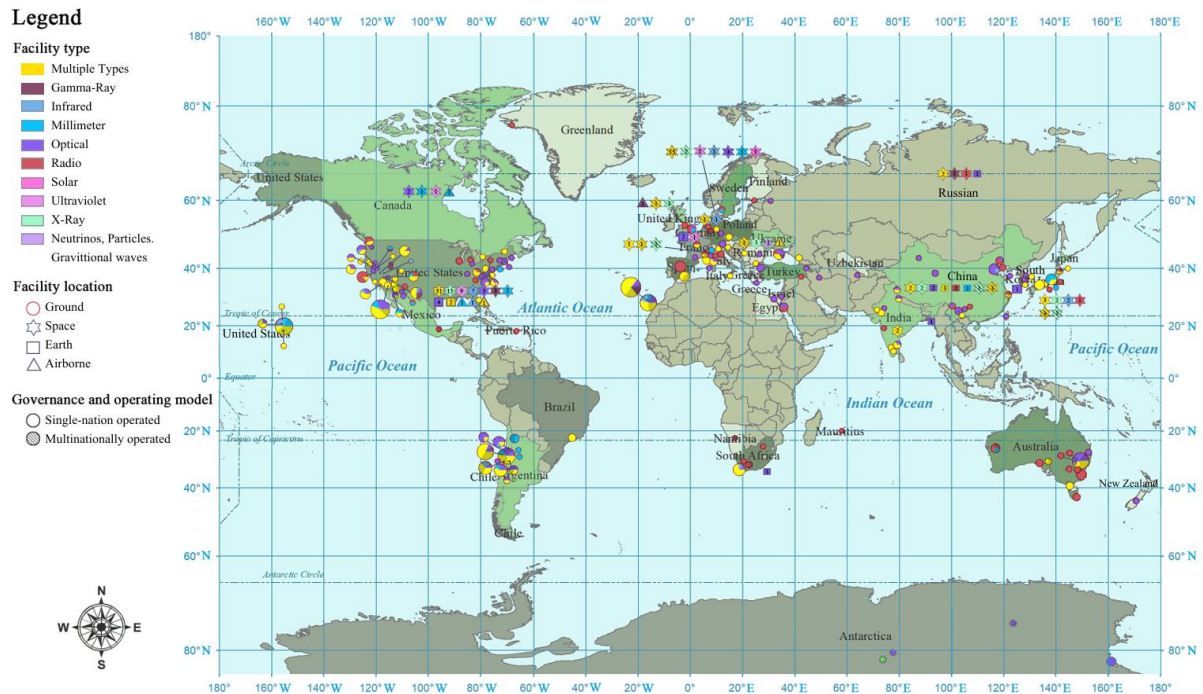

**Fig. 3** Global geography of 610 astronomical facilities (Ground facilities are mapped to their exact site coordinates. Space/Earth-orbiting/Airborne platforms are plotted at the operating country, for multinational consortia, the marker is assigned at random to one partner country)

A map of 610 facilities shows in Fig. 3, and three robust patterns emerge:

**Hemispheric asymmetry with clustered corridors.** Most installations lie in the Northern Hemisphere, concentrated in the western United States and Hawaii, Western/Central Europe (including the Canary Islands, Iberia, UK, France, Germany, Italy, Switzerland), and East Asia (Japan, China, Korea). In the Southern Hemisphere, three anchors dominate: the Atacama–Andes of Chile (optical/IR at Paranal/Cerro Pachón/Cerro Tololo and mm/sub-mm at Chajnantor), southern Africa (Sutherland/Karoo; SALT, MeerKAT, HERA; H.E.S.S. in Namibia), and Australia (Siding Spring, Parkes, ATCA, MRO/SKA-pathfinders), with additional nodes in New Zealand and Antarctica. Vast regions of equatorial Africa, interior Asia, and parts of SE Asia host few facilities, reflecting climate, infrastructure, and funding constraints.

**Site selection follows environmental/operational optima.** Optical/IR/mm facilities cluster at high, arid, and dark sites with stable seeing and low precipitable water vapor (Atacama, Mauna Kea, Canaries). Large radio arrays prefer radio-quiet, sparsely populated basins or deserts (Karoo, Murchison, New Mexico). Particle-astrophysics and gravitational-wave detectors occupy specialized environments—e.g., LIGO (US), Virgo (Italy), KAGRA (Japan) on seismically managed sites; IceCube at the South Pole for kilometer-scale Cherenkov volumes.

**Multi-wavelength co-location and network logic.** Numerous markers labeled "multiple types" indicate campus-style sites (e.g., Mauna Kea, La Palma/Tenerife, Paranal–Cerro Tololo–Pachón, Chajnantor) where shared logistics and atmosphere enable coordinated, multi-band observing. These sites emerge as infrastructural hubs that anchor collaborative constellations of institutions and enhance connectivity across spectral domains. When overlaid with the international collaboration network (Fig. 4), a structural correspondence becomes evident. Major facility clusters align with dense cross-national ties: the United States and Western Europe form the most central and interconnected hubs, with Japan and China emerging as secondary poles that reinforce global integration. Southern Hemisphere hosts such as Chile, South Africa, and Australia function as gateway nodes, coupling their unique geographic and atmospheric advantages to high levels of international participation. Meanwhile, Canada, the Nordic countries, and selected Central/Eastern European states leverage multinational facility access to insert themselves into otherwise core-dominated structures. From the dual perspective of facility co-location and national co-operation, global astronomy is not organized as a mere collection of isolated assets but rather as an interdependent system. Co-located sites provide the physical and technical substrate for coordinated observation, while the international collaboration network constitutes the social and organizational infrastructure that mobilizes these capacities. Together, they yield a distributed yet coherent architecture for all-sky coverage, multi-wavelength complementarity, and rapid follow-up in time-domain astronomy. This dual architecture also carries strategic significance, as it underpins the planning, sustainability, and global governance of next-generation large-scale facilities such as the SKA and ELT.

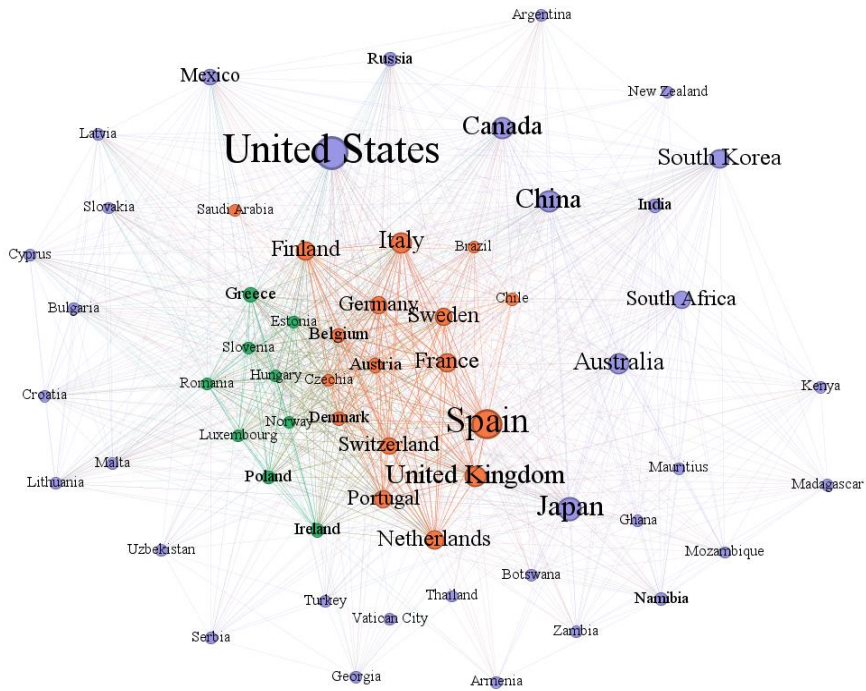

**Fig. 4** International co-operation network of facilities (60 country nodes, 1,013 links).

Fig. 3 and Fig. 4 reveal a globally connected but geographically concentrated infrastructure: a small number of premium environmental corridors host dense, multi-wavelength capacity, while international access policies and archival data diffuse scientific usage beyond host countries. This spatial structure underlies the leadership-participation patterns and inequality metrics analyzed in the subsequent sections.

By combining indicators of facility usage, scientific impact, and leadership intensity, we can characterize how different national hosts position themselves within the global observatory system. This multidimensional perspective goes beyond raw output, revealing structural asymmetries between infrastructure concentration, research participation, and leadership authority. These regularities are visible in Fig. 5. With these measures defined, several consistent patterns emerge:

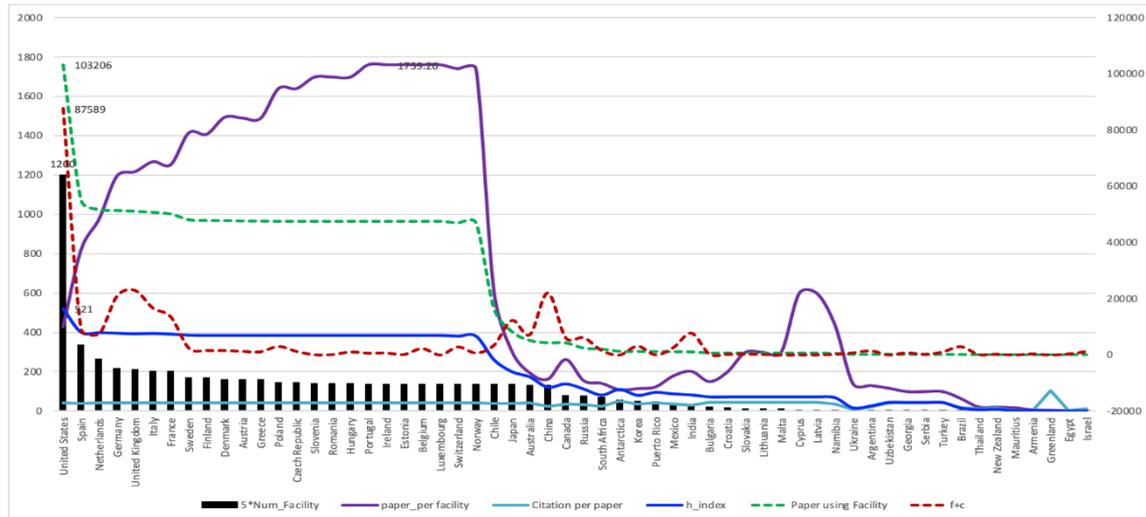

**Fig. 5** Facility stock, output, impact, per-facility productivity, and leadership intensity by country

Note: Black bars (left axis) show 5×Num_Facility (rescaling is for visualization only). The green dashed line (right axis) is total papers using domestic facilities. The blue line is the h-index of those papers (impact proxy), the cyan line the average citations per paper, the purple line the papers per facility (utilization), and the red dashed line (f+c) the sum of first- and corresponding-author papers (leadership intensity).

**Usage and impact are top-heavy.** A small set of hosts—the United States, leading Western European nations, China, Japan, and Australia—concentrate both usage (green) and impact (blue), confirming that high-impact work clusters in infrastructure hubs.

**Per-facility productivity is non-monotonic.** The purple series rises among several mid-scale European hosts (e.g., Netherlands, United Kingdom, Switzerland), indicating high utilization of compact, well-integrated portfolios, and then drops sharply as the facility base thins. Local spikes reflect niche efficiency but are sensitive to small denominators.

**Leadership partly decouples from volume.** The red dashed (f+c) curve shows peaks and troughs relative to usage. Observatory hosts such as Chile and South Africa exhibit lower leadership intensity than their usage would suggest (participation ≫ PI leadership), whereas some smaller systems show elevated leadership shares via focused programs and deep participation in multinational collaborations.

**Average citations per paper have limited discrimination beyond the leaders.** The cyan series is comparatively flat after the leading hubs, consistent with a shift from flagship programs to smaller projects and archival use; the h-index therefore provides the clearer impact signal at country scale.

Against this global backdrop, a regional perspective sharpens the picture and links facility portfolios, collaboration regimes, and PI capacity to distinct national profiles.

**North America (4 countries).** The United States leads on stock, usage, and impact and maintains high, stable utilization — an archetypal full-stack ecosystem (instrumentation, pipelines, PI programs). Canada outperforms its stock through intensive participation in multinational facilities. **Western Europe (ESO/ESA core) (14 countries)**. Network effects yield high volume and impact (Fig. 5). Netherlands, Switzerland, and the United Kingdom exhibit strong per-facility productivity, consistent with "few-but-excellent" portfolios. Leadership is robust but moderated by dense collaboration. **Nordics & Central/Eastern Europe (20 countries).** Nordic countries often achieve above-average utilization and respectable citation performance in specialized subfields. CEE countries form a steady long tail, where participation grows via European consortia despite modest domestic stock. **East/South/West Asia (10 countries).** China combines a growing stock with strong leadership intensity and rising impact (instrument- and PI-driven expansion); Japan shows balanced multi-band strength; Korea grows synchronously across series. India, Israel, and Turkey display niche efficiency or focused leadership despite smaller portfolios. **Oceania (2 countries).** Australia is balanced across stock–volume–impact with competitive per-facility productivity (radio/optical plus SKA-related programs). New Zealand shows episodic utilization spikes typical of small-N systems. **Latin America & Africa (7 countries).** Chile exemplifies the host-nation, high-participation pattern (high usage and impact with muted leadership). South Africa shows a similar structure (MeerKAT, SALT, HERA). Elsewhere, participation is primarily archive- and training-driven. **Special case: Antarctica.** Station activity is largely operational and multinational; leadership is typically registered abroad.

Country-level Gini coefficients corroborate the pattern: Num_Facility = 0.850 and Paper using Facility = 0.855 indicate strong head-concentration of infrastructure and the associated publication volume, while impact indicators are comparatively more diffuse (h_index = 0.805; Citation per paper = 0.713). A derived per-facility productivity measure has the lowest inequality (paper_per_facility = 0.518, computed for countries with ⩾1 facility), reflecting normalization by stock and the presence of efficient mid-scale hosts. Special cases such as Antarctica (operational, multinational bases) illustrate that site presence does not necessarily imply domestic PI output.

## 4.2 Who leads the research? Alignment between leadership and ownership

Leadership is more concentrated than either stock or usage. The Gini for first- and corresponding-author counts is near 0.91 (Author-f = 0.915; Author-c = 0.914), showing that opportunities to lead publications accrue to a small set of countries.

To connect these distributional results to their organizational underpinnings, we complement the Gini-based inequality profile with a structural view of how countries balance leading versus participating in publications. Fig. 6 links this distribution to national "roles". On the leadership‐participation map, each country deposited on a bivariate, log‐log map. The x-axis shows log (Author-f) (used here as a leadership proxy) and the y-axis log (Usage share) (participation proxy for papers where the country appears only as non-first, non-corresponding author). Dashed lines mark constant participation-to-leadership ratios (y=x, y=2x, y=0.75x). Symbol/label coding reflects facility location: red "×" with red label = host with ≥10 facilities; red "×" with black label = host with 1–9 facilities; black "×" with black label = no facilities. Countries with zeros cannot be plotted on log axes (128 country nodes in total).

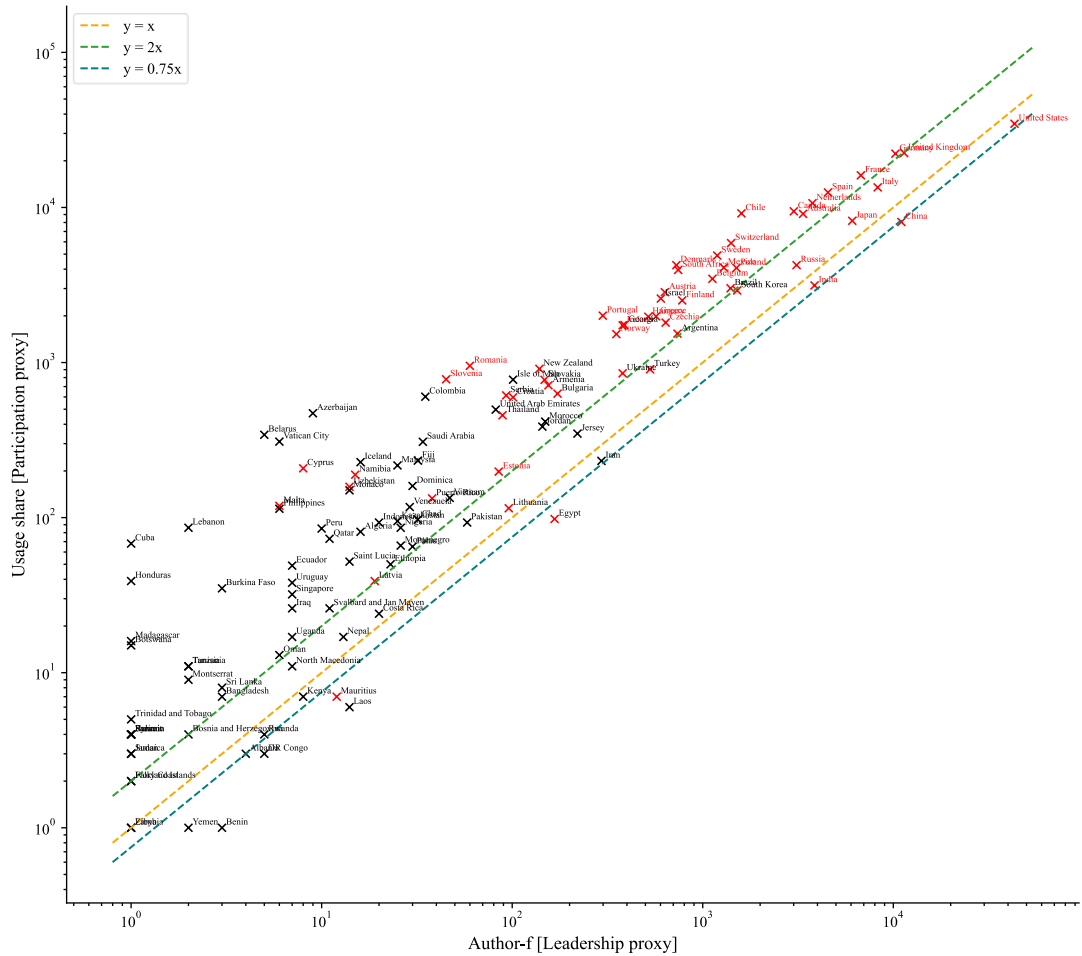

**Fig. 6** National Academic Positioning from Facility-Generated Publications: Leadership vs. Participation in Astronomy

Points follow an approximately linear trend (slope ≈ 1 in log–log space), indicating that participation and leadership scale together. Facility-rich hubs—United States, major Western European nations, China, Japan, Australia—cluster in the upper right near or slightly below y=x, consistent with strong leadership commensurate with participation (full-stack ecosystems). In contrast, Chile and South Africa lie well above y=x, exemplifying host-nation, high-participation systems where local facility density drives abundant co-authorship while much PI leadership resides abroad. Countries with few facilities (e.g., Mexico, New Zealand, Israel, Brazil, India) occupy intermediate regimes—access raises participation, but limited platform scale constrains aggregate leadership. Nations without facilities cluster lower left yet remain present through archives and international consortia, underscoring the equalizing role of open data; durable rightward movement typically requires domestic PI opportunities, instrument/processing capacity, and stable funding.

Facility–leadership decoupling. A notable phenomenon is facility-rich yet "leadership-silent" locations: Luxembourg (27 facilities), Antarctica (11), and Greenland (1) appear in the facility inventory but yield zero corresponding-author papers in our window, so they cannot be placed on log axes. This likely reflects ground-segment/infrastructure roles (tracking, relay, geodetic) and multinational bases where PI affiliations and authorship credit are registered abroad, as well as the temporal/domain coverage of our bibliometrics. The case underscores that operational presence does not automatically translate into domestic PI leadership.

Synthesis. Read jointly Fig. 4 and Fig. 6, stock (black), usage (green), impact (blue/cyan), efficiency (purple), leadership (red), and the leadership–participation map reveal three regimes:

(1) Infrastructure hubs with high output, high impact, and durable leadership;

(2) Efficient mid-scale hosts that convert compact portfolios into high per-facility yield (and sometimes strong leadership);

(3) Low-infrastructure participants with modest absolute output and weaker unit productivity. Facilities lay the foundation for volume and impact, but domestic PI programs, instrumentation/data pipelines, and collaborative integration determine how effectively each facility is converted into leadership and high-value papers.

## 4.3 Collaboration structure and power network

To understand the distribution of collaboration and authority in global astronomy, we analyze the power network from two complementary perspectives. First, the author–institution–country network (Fig. 7) captures the structural backbone of collaboration, highlighting which countries, institutions, and individual researchers occupy central bridging positions within the global scientific community. Second, the facility–first author–corresponding author network (Fig. 8) reveals the functional roles embedded in these collaborations, distinguishing between sites of research execution and sources of leadership authority. Taken together, these two layers of analysis provide a comprehensive view of how scientific power is simultaneously structured and enacted: structurally through positions in the collaboration network, and functionally through the division of labor between active research participation and project leadership.

## 4.3.1 Multi-layer collaboration networks (Author–Institution–Country)

The three-layer collaboration network (authors–institutions–countries) reveals a hierarchical organization of scientific cooperation within global astronomical research (Fig. 7).

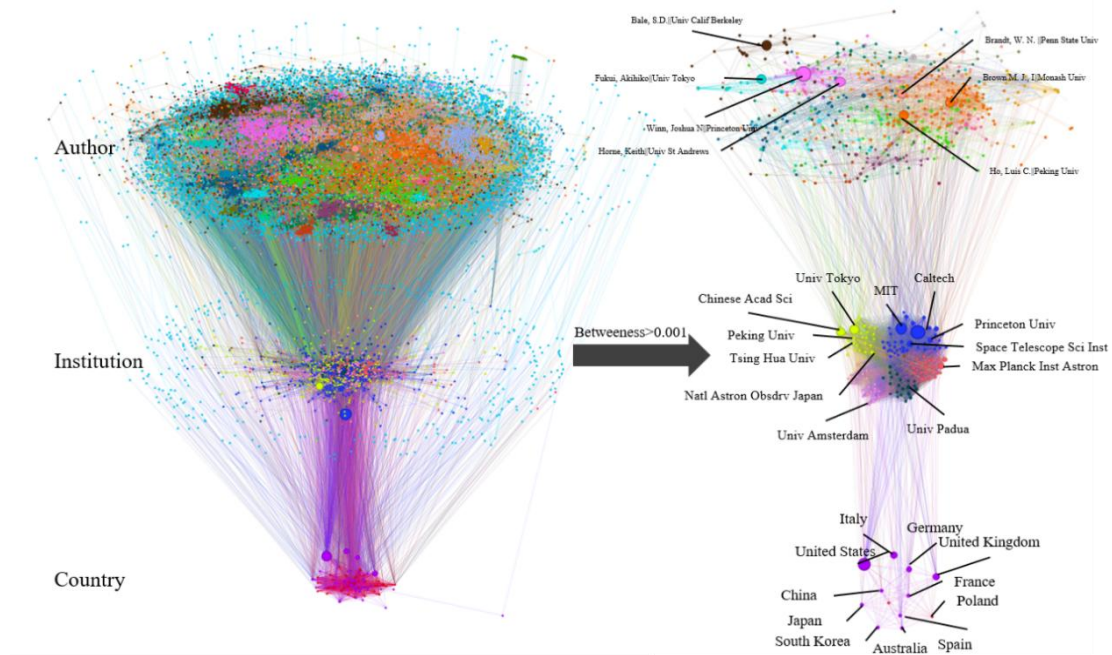

**Fig. 7** Global network of Author–Institution–Country (2020-2025)

**National level.** The United States emerges as the dominant hub, bridging multiple international clusters and mediating the largest share of global collaborations. China, Germany, the United Kingdom, and Japan also hold high betweenness centrality, underscoring their dual role as both major producers of astronomical research and indispensable intermediaries in the international knowledge network. Notably, China has moved beyond peripheral participation and, through its top institutions, has consolidated a visible position in the global collaboration landscape.

**Institutional level.** A small set of universities and research institutes function as structural cores of the network. Institutions such as MIT, Caltech, the University of Tokyo, the Chinese Academy of Sciences, Peking University, Tsinghua University, Princeton University, the Max Planck Institute for Astronomy, and the University of Padua are situated at the crossroads of transnational collaborations. Their high betweenness centrality indicates that they are not only prolific in scientific output but also act as "brokers" linking otherwise disconnected regional research communities.

**Author level.** A handful of individual researchers, such as Fukui (University of Tokyo), Brandt (Pennsylvania State University), Brown (Monash University), and Winn (Princeton University), appear as bridging nodes between distinct institutional clusters. Their positions demonstrate how individual scientists act as micro-level connectors, enabling knowledge to circulate across institutional and national boundaries.

Taken together, the three-layer perspective shows how power and connectivity are simultaneously concentrated and distributed: concentrated in a few leading countries and institutions, yet distributed through a broader web of individual researchers. This layered structure emphasizes that dominance in the global astronomy community is not only determined by research output or facility ownership but also by the ability to occupy central bridging positions across multiple levels of collaboration. Specific details regarding the central actors at each level are summarized in Table 3.

**Table 3** Details regarding the central actors at each level

| No. | Author | Betweeness | No. | Institution | Betweeness | No. | Country | Betweeness |
|---|---|---|---|---|---|---|---|---|
| 1 | Winn, Joshua N.\|\|Princeton Univ | 0.283106473 | 1 | CALTECH | 0.456583099 | 1 | United States | 0.715490566 |
| 2 | Seager, S.\|\|Massachusetts Institute of Technology | 0.24470797 | 2 | Massachusetts Institute of Technology | 0.19461834 | 2 | Italy | 0.101426998 |
| 3 | Steller, M.\|\|Austrian Acad Sci | 0.077239525 | 3 | Univ Tokyo | 0.157798619 | 3 | United Kingdom | 0.09663146 |
| 4 | Horne, Keith\|\|Univ St Andrews | 0.069809629 | 4 | Chinese Acad Sci | 0.140251842 | 4 | Germany | 0.070411961 |
| 5 | Quirrenbach, A.\|\|Heidelberg Univ | 0.067864872 | 5 | Space Telescope Sci Inst | 0.080963922 | 5 | China | 0.028887696 |
| 6 | Bale, S. D.\|\|Univ Calif Berkeley | 0.067008647 | 6 | Max Planck Inst Astron | 0.07943372 | 6 | France | 0.026911706 |
| 7 | Fukui, Akihiko\|\|Univ Tokyo | 0.056673732 | 7 | Univ Calif Berkeley | 0.050554866 | 7 | Japan | 0.026529812 |
| 8 | Pollacco, D.\|\|Univ Warwick | 0.056567619 | 8 | Princeton Univ | 0.047491865 | 8 | South Korea | 0.026118326 |
| 9 | Brandt, W. N.\|\|Penn State Univ | 0.055716179 | 9 | Univ Geneva | 0.045734926 | 9 | Australia | 0.026114356 |
| 10 | Trump, Jonathan R.\|\|Univ Connecticut | 0.053594236 | 10 | Leiden Univ | 0.03344225 | 10 | Spain | 0.022739535 |
| 11 | Ho, Luis C.\|\|Peking Univ | 0.053140444 | 11 | Natl Astron Observ Japan | 0.032774899 | 11 | Poland | 0.00103397 |
| 12 | Palle, E.\|\|Inst Astrofis Canarias | 0.052967564 | 12 | Univ Arizona | 0.032717766 | 12 | Russia | 0.000986317 |
| 13 | Henning, Th.\|\|Max Planck Inst Astron | 0.052436207 | 13 | Aix Marseille Univ | 0.031079485 | 13 | Netherlands | 0.000595916 |
| 14 | Santos, N. C.\|\|Univ Porto | 0.052142522 | 14 | NASA | 0.03104713 | 14 | Romania | 0.000595916 |
| 15 | Bonfils, X.\|\|Univ Grenoble Alpes | 0.047604105 | 15 | Univ Padua | 0.029996511 | 15 | Brazil | 0.000595916 |
| 16 | Koekemoer, Anton M.\|\|Space Telescope Sci Inst | 0.04377351 | 16 | Univ Grenoble Alpes | 0.0274744 | 16 | Czechia | 0.000555786 |
| 17 | Stappers, B. W.\|\|Univ Manchester | 0.042864138 | 17 | Univ Amsterdam | 0.026559566 | 17 | Switzerland | 0.000341109 |
| 18 | Merloni, A.\|\|Max Planck Inst Extraterr Phys | 0.041099005 | 18 | Univ Michigan | 0.026484597 | 18 | Georgia | 0.000333996 |
| 19 | Breton, R. P.\|\|Univ Manchester | 0.039759535 | 19 | Stanford Univ | 0.024700797 | 19 | Norway | 0.000243045 |
| 20 | Udry, S.\|\|Univ Geneva | 0.038916379 | 20 | Ohio State Univ | 0.023485426 | 20 | Slovakia | 0.000243045 |
| 21 | Cha, Sang-Mok\|\|Kyung Hee Univ | 0.034626279 | 21 | Penn State Univ | 0.022910616 | 21 | South Africa | 0.000243045 |
| 22 | Fan, Xiaohui\|\|Univ Arizona | 0.032820082 | 22 | Univ Warwick | 0.021888992 | 22 | Mexico | 0.000243045 |
| 23 | Stern, Daniel\|\|CALTECH | 0.031906693 | 23 | Univ Cambridge | 0.021310458 | 23 | India | 0.0001443 |
| 24 | Leroy, Adam K.\|\|Ohio State Univ | 0.031180911 | 24 | Univ Calif Santa Cruz | 0.020047326 | 24 | Denmark | 0.00011614 |
| 25 | Gromadzki, M.\|\|Univ Warsaw | 0.03015918 | 25 | Heidelberg Univ | 0.015281162 | 25 | Sweden | 0.00011614 |
| 26 | Gromadzki, Mariusz\|\|Univ Warsaw | 0.02967572 | 26 | Polish Acad Sci | 0.014948912 | 26 | Turkey | 0.00011614 |
| 27 | Stassun, Keivan G.\|\|Vanderbilt Univ | 0.029630593 | 27 | Univ Maryland | 0.014833639 | 27 | Greece | 0.00011614 |
| 28 | Beuther, H.\|\|Max Planck Inst Astron | 0.02788025 | 28 | Univ Turku | 0.014548359 | 28 | Ukraine | 0.000114319 |
| 29 | Carretero, J.\|\|Barcelona Inst Sci & Technol | 0.026660438 | 29 | Kyoto Univ | 0.013716082 | 29 | Colombia | 9.77036E-05 |
| 30 | Graham, Matthew J.\|\|CALTECH | 0.026489694 | 30 | Natl Radio Astron Observ | 0.013414757 | 30 | Slovenia | 3.41763E-05 |

## 4.3.2 Facility-based roles in execution and leadership (Facility–First Author–Corresponding Author)

Fig.8 provides a complementary perspective by linking facility locations, first author affiliations, and corresponding author affiliations through a three-layer Sankey diagram. Each layer captures a distinct dimension of research contribution and authority.

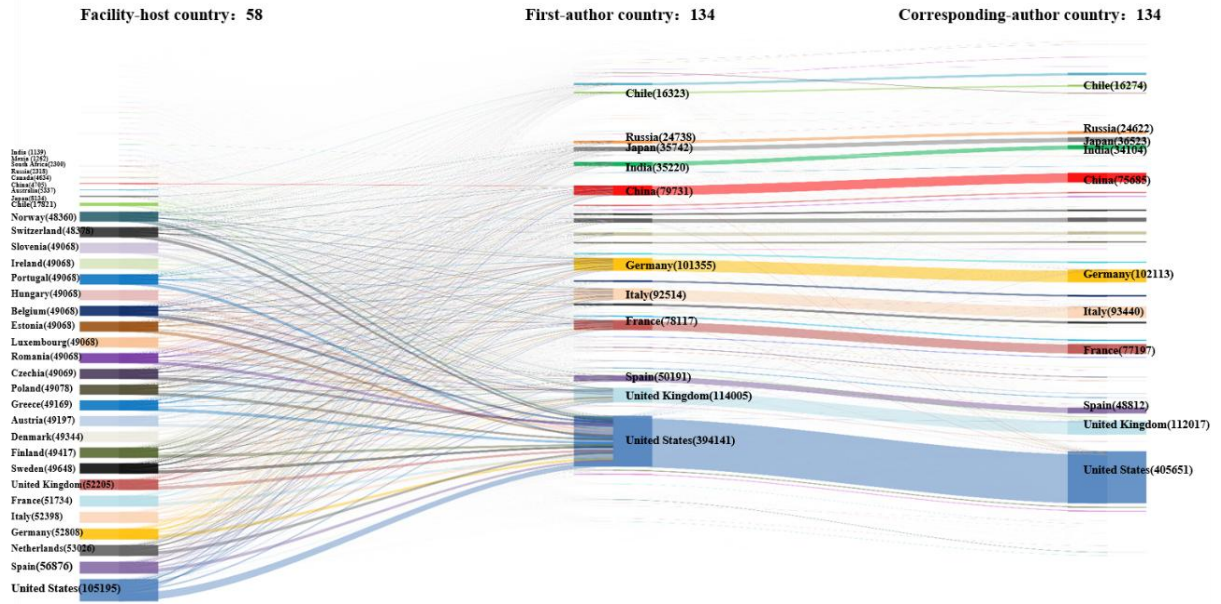

**Fig. 8** Network linking facility location, first author affiliation, and corresponding author affiliation

The facility country identifies where the enabling infrastructure is located, reflecting which nations provide the material basis of observation and experimentation. The first author's country generally signals the site of active research execution, since first authors are most often doctoral candidates, postdoctoral researchers, or early-career scientists who directly conduct analyses and prepare manuscripts. This dimension therefore reflects the distribution of "scientific labor and activity" across countries, highlighting where the most active strata of the research workforce are embedded. By contrast, the corresponding author's country reflects seniority, project leadership, and resource ownership, as corresponding authors are typically group leaders or principal investigators responsible for securing funding, guiding research design, and assuming accountability for results. Corresponding authorship thus represents "scientific authority and power."

Viewed through this layered lens, the Sankey diagram illustrates a differentiated global pattern of scientific roles. The United States dominates both execution and leadership, forming a closed domestic cycle of facilities, first authors, and corresponding authors. It also emerges as the principal sink for leadership internationally, frequently occupying corresponding-author positions even when facilities are hosted abroad (e.g., in Spain, Germany, Italy, the United Kingdom, France, and Chile). This asymmetry highlights how scientific authority remains concentrated in U.S. institutions, while the underlying infrastructure may be geographically dispersed.

European countries (Germany, Italy, France, the United Kingdom, Spain) exhibit a more balanced structure. Although their facilities feed substantial flows into U.S. leadership, they also sustain considerable domestic leadership at the corresponding-author stage. Their systems therefore appear more pluralistic, with domestic researchers actively engaged not only in execution but also in project direction.

Despite strong outflows to the U.S., European science producers (Germany, Italy, France, the United Kingdom, Spain) still retain significant domestic leadership, particularly at the corresponding-author stage. Their facilities produce many nationally led publications, although the scale is more distributed than the concentrated U.S. hub.

Asian countries (China, Japan, India) reveal a more nationally self-contained cycle: first and corresponding authors are predominantly from the same country, pointing to strong domestic pipelines of execution and leadership but relatively limited penetration into global leadership roles.

Chile represents a distinct case. Despite hosting world-class astronomical facilities, it contributes relatively few first or corresponding authors. This positions Chile primarily as a provider of infrastructure, while scientific execution and leadership are disproportionately captured by U.S. and European researchers.

Taken together, the Sankey diagram demonstrates that execution and authority are unevenly distributed. Countries with many first authors can be understood as hubs of research activity, with large cohorts of active scientists performing day-to-day work. By contrast, countries with many corresponding authors concentrate scientific leadership and decision-making power, shaping the strategic direction of facility-based astronomy. This distinction underscores that dominance is not merely a function of data production, but also of the ability to direct and own research through authorship hierarchies.

### 4.4 Evolution of leadership

### 4.4.1 Rebalancing of lead authorship

The change trends of each country's share of papers with a domestic first author (Fig. 9) shows a gradual redistribution of leadership from a single dominant pole to a small set of hubs. The U.S. remains the single largest source of first authors, but its global share shows a persistent decline from ~40% in the early 2000s to ~25% in 2024 (fitted slope ≈ −0.0071 per year). This indicates a gradual redistribution of first-authorship leadership away from a once-dominant pole. China's trajectory is the mirror image, a steady and accelerating rise from near zero at the start of the period to ~16% by 2024 (slope ≈ +0.0054 per year), with visible acceleration after 2015. The gap with the U.S. narrows materially, though no crossover occurs within the window. The United Kingdom, Germany, France, Italy, the Netherlands, and Japan occupy the 3–10% band. Most show flat to gently declining shares, consistent with the diffusion of leadership across a widening global base and intra-European dispersion across multiple countries and consortia. India trends upward (toward ~5%), while several smaller systems (e.g., Chile, Spain) show low but sometimes volatile shares, reflecting small-N sensitivity.

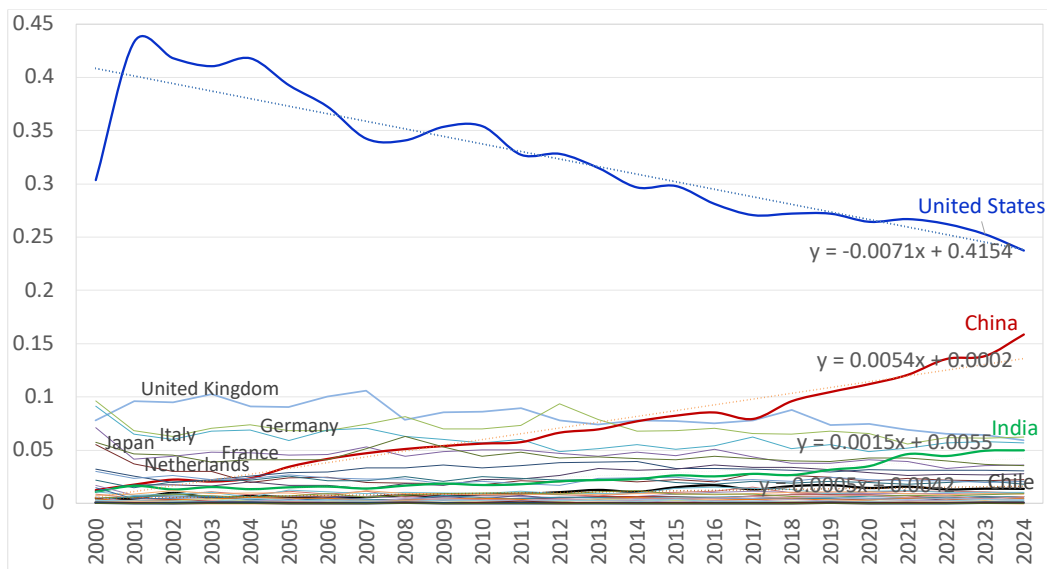

**Fig. 9** Annual first-author shares by country located with more than 10 facilities (2000–2024)

The panel points to an ongoing rebalancing of authorship leadership. the U.S. remains first by a clear margin, China is converging quickly, and the second tier of European and Asian producers is stable to slightly eroding in relative terms. Because the figure reports shares, the U.S. decline does not imply falling absolute output; rather, it reflects faster growth elsewhere. Short-term wiggles for small countries chiefly reflect denominator effects; the linear fits summarize the directional trend in leadership shares over the period.

**4.4.2 Inequality dynamics**

The change trend of the country-level Gini coefficients of the share of papers with domestic first authors (Author-f, blue) and corresponding authors (Author-c, red) shows the inequality dynamics. Authorship leadership has become more centralized among a small set of countries (Fig. 10). Both series increase from about 0.82 in 2000 to ≈0.88 by 2024. The fitted linear trend (blue, y=0.0022x+0.823) implies an average rise of ~0.0022 per year (≈+0.05 over 24 years). First- and corresponding-author leadership move in lockstep. The blue and red curves track each other closely with only minor divergences, indicating that credit allocation for first and corresponding authorship follows the same geopolitical structure.

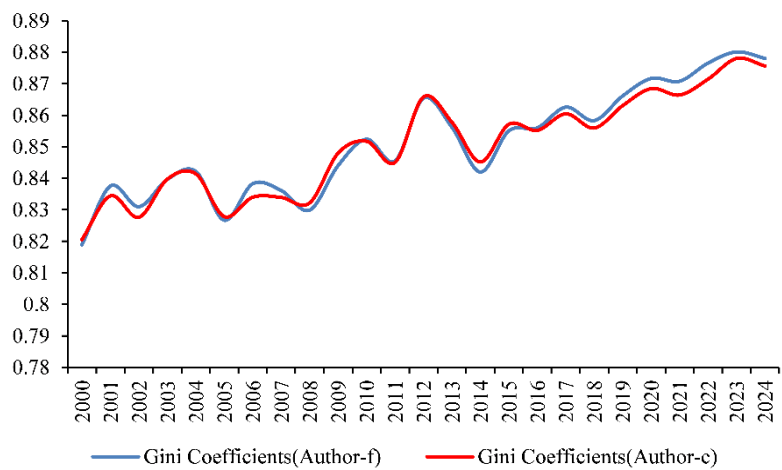

**Fig. 10** Gini coefficients for first- and corresponding-author shares over 2000-2004

Despite the broadening of global participation, lead authorship has not diversified proportionally. The declining U.S. share and the strong rise of China do not reduce inequality; instead, leadership becomes concentrated in a handful of hubs (U.S., major Western European nations, China, Japan, Australia), while many other countries remain in a long tail with small or zero leadership shares. Large survey collaborations and centralized PI structures likely reinforce this pattern.

**4.4.3 Infrastructure and its timing**

The fraction changes of global facility-based papers attributable to facilities hosted in that country (Fig. 11) reveal when and where facilities underwrote this shift. United States dominates throughout. its host share is near 100% when the modern ground-based system is nascent (late 1950s–60s), then falls to a broad 70–80% plateau in the 1970s–2000s and declines further to 68% in the 2020s—evidence of gradual diversification away from a single pole rather than an absolute contraction of U.S. output.

A tightly co-moving block of Western European hosts—Spain, Netherlands, United Kingdom, Germany, France, Italy—rises steeply from the late 1980s through the 1990s, coincident with the maturation of ESO/ING/La Palma and other European infrastructures, peaks around the late-1990s/early-2000s, softens in the mid-2000s, and recovers in the 2010s–2020s. The near-parallel trajectories indicate a coupled facility system driven by common consortia and time-allocation mechanisms.

Japan shows an early rise with a mid-1990s crest, consistent with the commissioning of large optical/IR facilities and national programs, followed by a leveling off. Chile exhibits a monotonic increase from ~2005 onward, reflecting its role as the premier Southern Hemisphere host (VLT/Paranal, Cerro Pachón/Tololo, and ALMA). The curve's climb indicates growing reliance on Chile-hosted sites for global publications. China emerges from near zero in the 1990s to a visible and accelerating host share after ~2010–2015, aligning with the ramp-up of LAMOST, FAST, HXMT and expanded domestic observatory capacity.

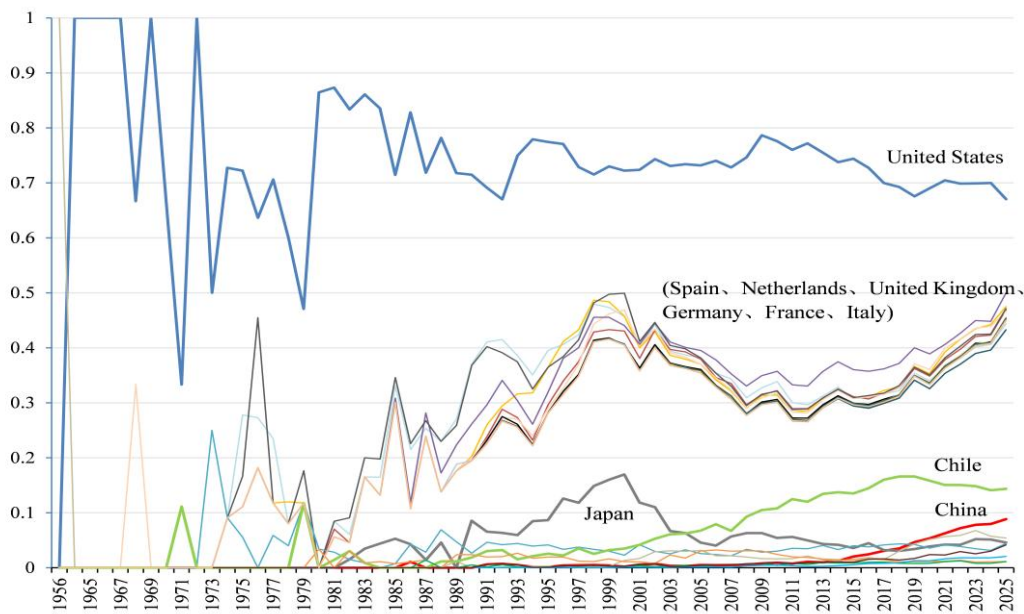

**Fig. 11** host-country shares of papers that use domestic facilities (1955–2025)

The figure captures a rebalancing of host-country contributions: the U.S. remains the primary hub but its relative dominance erodes as European, Southern Hemisphere, and Chinese infrastructures expand. The lockstep movement among Western European curves reveals integration via shared facilities and consortia, while Chile's steady rise reflects the consolidation of high-altitude, low-PWV sites for optical/IR and mm/sub-mm astronomy. Japan's earlier peak and China's later acceleration illustrate different timing in facility build-out.

It is should be noted that early-period fluctuations reflect small-N volatility. "Share" is by host location of the facility (not PI nationality) and therefore highlights where infrastructure is sited, not where scientific leadership resides—a distinction explored in the leadership/participation analysis that follows.

### 4.4.4 Impact profile of hosted facilities

Each curve gives the yearly h-index of the cohort of papers that used facilities hosted in a given country and were published in that year (i.e., not cumulative over all prior years) in Fig. 12. Higher values indicate that more papers from that year's cohort received at least that many citations, combining volume and impact. The system exhibits a long "infrastructure wave": a rapid upswing from the late-1980s through the 2000s, a broad plateau around the early 2010s, and a decline into the 2020s. The drop in the right tail is expected because recent cohorts have had less time to accrue citations (citation-window/lateness effect), with the pandemic years adding further headwinds.

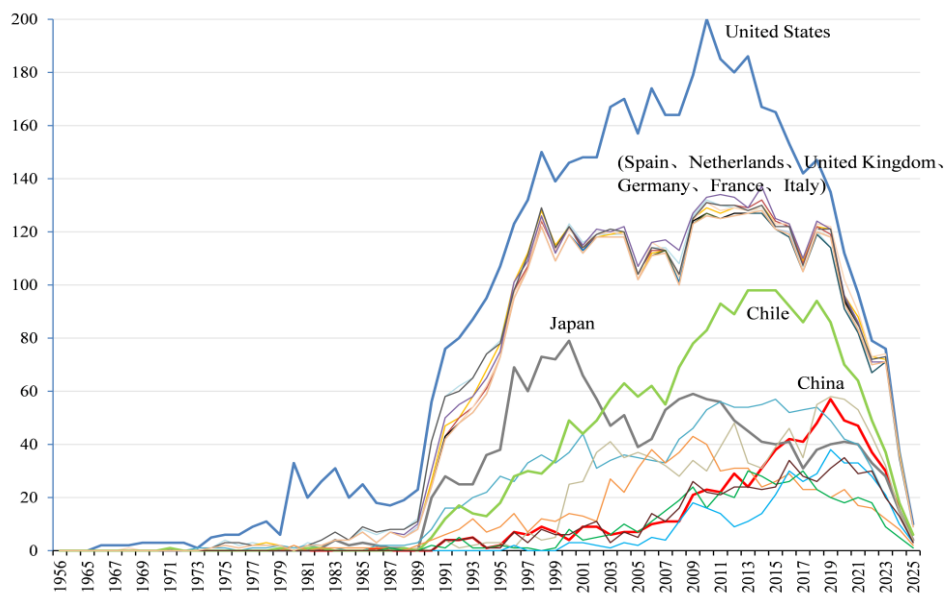

**Fig. 12** The yearly h-index of papers produced with facilities sited in each country (1955–2025).

The U.S. curve dominates throughout, surging in the late-1980s/1990s, peaking near ~200 in the early 2010s, and easing thereafter. This reflects the scale and cadence of U.S. facility programs (HST/Keck/SDSS/Chandra/Fermi, etc.) and their large collaborative ecosystems. Western Europe (ESO/ESA–ING core: Spain, Netherlands, United Kingdom, Germany, France, Italy), these countries move nearly in lockstep, rising steeply in the 1990s with the maturation of ESO/ING/Canaries and Europe-led space missions, reaching ~100–130 around the late-2000s/early-2010s, dipping mid-2010s, and recovering slightly thereafter. The tight parallelism indicates high integration via shared facilities and consortia.

Japan shows an earlier crest (late-1990s) associated with national optical/IR and X-ray programs (e.g., ASCA, Subaru era), followed by a modest decline/leveling—consistent with program turnover and global scaling elsewhere. Chile's h-index rises monotonically from the mid-2000s, cresting near ~90–100 in the mid-2010s. This is the signature of its role as the premier Southern Hemisphere host (VLT/Paranal, Cerro Pachón/Tololo, ALMA/Chajnantor), where globally led projects produce high-impact cohorts at Chile-hosted sites. China transitions from near zero to a steep ascent after ~2010, reaching ~50+ before the right-tail decline. The timing aligns with the ramp-up of LAMOST, FAST, HXMT and expanded domestic platforms—evidence of rapidly rising impact of China-hosted facilities.

The panel indicates a redistribution of facility-based impact: the U.S. remains first but its margin narrows as European, Chilean, and Chinese infrastructure waves mature. The synchronous movement of the European block reveals coordinated capacity building; Chile shows how hosting world-class sites translates into high-impact cohorts even when PI leadership often resides abroad (as seen in authorship analyses); Japan and China illustrate different timing of national build-outs.

Early-period volatility (1950s–1970s) reflects small-N sensitivity. Because the h-index here is cohort-year based, declines in the most recent years should not be read as structural deterioration but as citation-window effects; multi-year rolling windows would smooth those edges.

## 5. Conclusions

This study provides a comprehensive examination of the global geography, usage,

and leadership of astronomical facilities, situating infrastructures within the broader dynamics of research participation, impact, and collaboration. Several key conclusions emerge.

First, astronomy remains shaped by a pronounced spatial asymmetry. Facilities are geographically concentrated in a few premium environmental corridors in the Northern and Southern Hemispheres, yet their usage is internationally diversified through time-allocation systems, archival data, and multinational consortia. This dual structure—geographical concentration coupled with broad scientific access—anchors the distributed but interdependent character of global astronomy.

Second, scientific leadership is significantly more concentrated than facility hosting or usage. A handful of countries—above all the United States, followed by leading European nations, China, Japan, and Australia—account for the majority of first- and corresponding-author positions. Host nations such as Chile and South Africa exemplify the "infrastructure–participation" decoupling: while they underpin a large fraction of global observational capacity, much of the leadership credit accrues abroad. This reveals a structural inequality in which facility provision does not automatically translate into research authority.

Third, collaboration networks display a layered and hierarchical organization. At the macro level, a small number of countries and institutions occupy bridging positions that channel the flow of knowledge. At the meso level, leading universities and institutes function as structural cores, connecting otherwise fragmented regions. At the micro level, key authors emerge as brokers, linking different institutional clusters. Complementary analysis of facility–first–corresponding author flows demonstrates the functional division of labor: first authorship reflects scientific labor and activity, while corresponding authorship embodies authority and control. Together, these perspectives highlight how power in astronomy is both concentrated in dominant hubs and distributed through a broader web of collaborators.

Fourth, leadership dynamics are evolving. The United States remains the dominant hub but has experienced a gradual decline in its global share of first authorship, even as absolute output continues to grow. China has rapidly ascended to become the second-largest source of lead authors, reflecting major investments in domestic facilities and talent pipelines. Europe's contribution remains strong but internally distributed across multiple national systems. The trajectory indicates a slow rebalancing of authorship leadership from a single dominant pole toward a multipolar set of hubs, even as inequality metrics show continued concentration of power.

Finally, the temporal analysis of impact underscores the "infrastructure wave." Successive phases of facility build-out—in the United States, Europe, Japan, Chile, and more recently China—are clearly reflected in the h-index trajectories of facility-based papers. These waves illustrate how long-term infrastructure investments translate into globally visible scientific impact, but also how the timing and integration of facilities determine their leverage in shaping leadership and collaboration outcomes.

Taken together, these findings underscore that global astronomy is not simply organized by where facilities are located, but by how infrastructures are embedded in collaborative regimes, leadership hierarchies, and evolving geopolitical dynamics. The asymmetries documented here carry implications for the design and governance of next-generation facilities such as the Square Kilometer Array (SKA) and the Extremely Large Telescope (ELT). Ensuring equitable participation, balancing leadership opportunities, and fostering inclusive governance will be central to sustaining astronomy as a truly global enterprise.